# Limits of topological protection under local periodic driving

Z. Cherpakova,[1,*] C. Jörg,[2,*] C. Dauer,[2] F. Letscher,[2,3] M.Fleischhauer,[2] S. Eggert,[2] S. Linden,[1,+] and G. von Freymann[2,4,±]

[1]*PhysikalischesInstitut, Universität Bonn, 53115 Bonn, Germany*

[2]*Physics Department and Research Center OPTIMAS, TU Kaiserslautern, 67663 Kaiserslautern, Germany*

[3]*Graduate School Materials Science in Mainz, 67663 Kaiserslautern, Germany*

[4]*Fraunhofer Institute for Industrial Mathematics ITWM, 67663 Kaiserslautern, Germany*

*these authors contributed equally to this work

\+ linden@physik.uni-bonn.de

± georg.freymann@physik.uni-kl.de

The bulk-edge correspondence guarantees that the interface between two topologically distinct insulators supports at least one topological edge state that is robust against static perturbations. Here, we address the question of how dynamic perturbations of the interface affect the robustness of edge states. We illuminate the limits of topological protection for Floquet systems in the special case of a static bulk. We use two independent dynamic quantum simulators based on coupled plasmonic and dielectric photonic waveguides to implement the topological Su-Schriefer-Heeger model with convenient control of the full space- and time-dependence of the Hamiltonian. Local time periodic driving of the interface does not change the topological character of the system but nonetheless leads to dramatic changes of the edge state, which becomes rapidly depopulated in a certain frequency window. A theoretical Floquet analysis shows that the coupling of Floquet replicas to the bulk bands is responsible for this effect. Additionally, we determine the depopulation rate of the edge state and compare it to numerical simulations.

## I. INTRODUCTION

In recent years, topology has been developed into a powerful concept to classify condensed matter systems beyond the Landau paradigm of spontaneous symmetry breaking. One of the important findings is that the topological properties of the bulk Hamiltonian can have a profound impact on the character of the modes at the boundary of the system. According to this bulk-boundary correspondence principle, the interface between two insulators with different topologies supports at least one conducting edge state that is protected by topology, *i.e.*, it supports a current along the interface without scattering even in the presence of strong static deformations [1, 2]. This intriguing property has been observed in a number of solid-state [3, 4], photonic [5] and cold atom systems [6].

A powerful tool for manipulating various quantum systems is time-periodic driving. The underlying principle is that driving of a system with frequency $\omega$ enables the hybridization of eigenstates of a static system, which are separated in energy by a multiple of $\hbar\omega$. As a result, new synthetically designed properties, inaccessible in equilibrium, can emerge. For instance, appropriately chosen driving regimes allow for coherent control of single-particle tunnelling [7], tuning transport regimes from ballistic to localized [8, 9], and inducing quantum phase transitions [10]. In addition to the driving frequency and amplitude, the spatial extent of the driving is also a valuable degree of freedom. As an example, by periodically driving individual lattice sites, one can control the transmission across the modulated region [11–14], pump charge [15], and create new Floquet bound states [14].

Periodic driving can change the topological properties of a system. In particular, a system, trivial in equilibrium, can become a topological insulator under periodic driving [16–18]. In systems with time-periodic driving, the bulk-edge correspondence needs to be generalized, and anomalous edge modes can exist [19, 20]. Time-periodic disorder at the boundary can also induce a shift in the energy of the topological edge state under certain conditions [21].

While in the static case, the coupling of the edge state to bulk states is energetically forbidden, dynamic perturbations of the system might result in hybridization of the modes and drastically change their character. Hence, it is important to understand under which conditions such a hybridization becomes relevant and when not. In this paper, we combine two different dynamic quantum simulators based on both plasmonic and dielectric coupled waveguides (see Fig. 1) together with a full Floquet theoretical analysis in order to study the characteristics of topological protected edge states under local time-periodic driving. Applying perturbations locally to the edge while keeping the bulk static allows us to study the limits of topological protection for special Floquet systems. We analyse such perturbations in the Su-Schriefer-Heeger (SSH) model, a simple yet topologically non-trivial system. The unique combination of two independent experimental quantum simulators allows for precise control of the system's parameters as well as an uncomplicated detection technique [22–26].

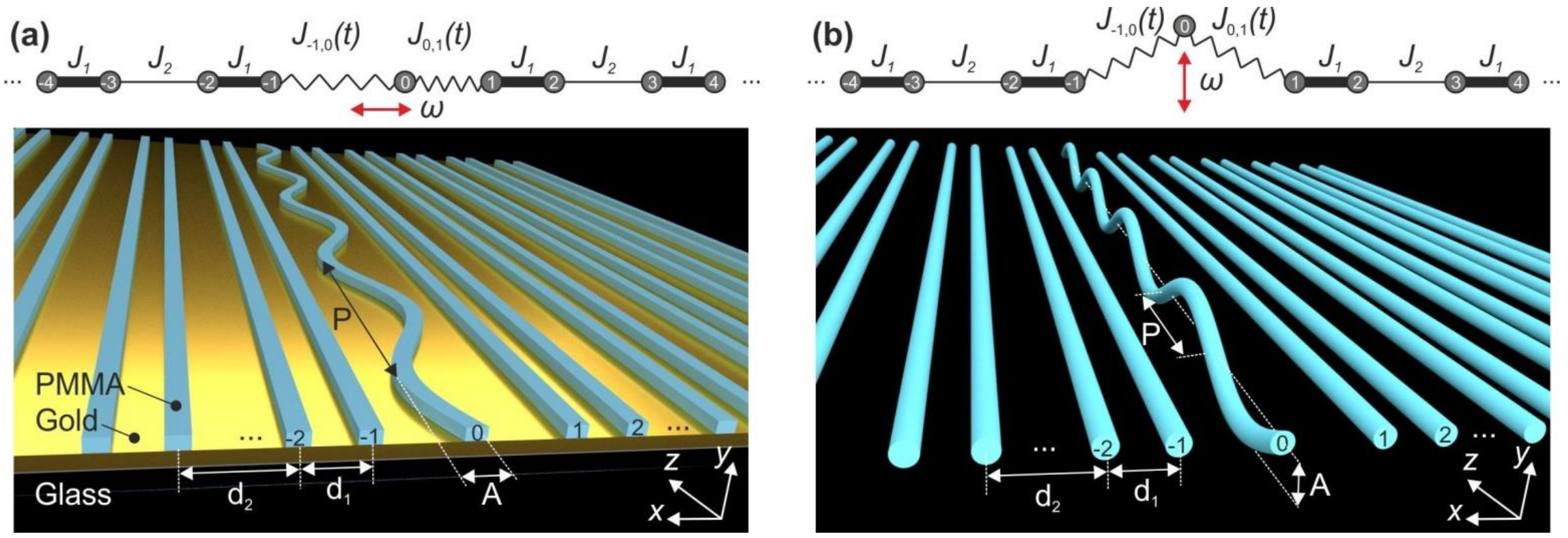

FIG. 1. Sketches of the SSH chains with time-periodic perturbations of a single lattice site at the interface between two distinct dimerizations (top) and the corresponding experimental realizations (bottom). (a) In-plane modulation of the boundary implemented in a plasmonic waveguide array. (b) Out-of-plane modulation of the boundary implemented in a dielectric waveguide array. Here, $J_1$ ($J_2$) denotes the large (small) hopping amplitude in the bulk, $J_{0,1}(t)$ ($J_{-1,0}(t)$) is the periodically modulated hopping amplitude between the $0^{th}$ and $1^{st}$ lattice sites ($0^{th}$ and $-1^{st}$ lattice sites), $\omega$ is the driving frequency, $d_1$ ($d_2$) is the short (long) centre-to-centre distance, $A$ is the maximum deflection of the $0^{th}$ waveguide from the centre, and $P$ is the period of driving. In the out-of-plane modulation, there is no difference if the waveguide bends up or down, and hence, the on-site potential and couplings vary with twice the waveguide period. Therefore, we define the period of modulation $P$ to be half the waveguide period in (b). Note that in the waveguide system, the propagation distance $z$ corresponds to time $t$.

In the static case, the SSH model describes a chain of identical lattice sites with alternating strong and weak bonds [27] (denoted here as $J_1$ and $J_2$, respectively) that can be implemented by alternating short $d_1$ and long $d_2$ distances between adjacent waveguides, respectively. Depending on the choice of the unit cell, the SSH model exhibits two topologically distinct dimerizations [28]. At each interface between two domains of different topologies, a topologically protected edge state occurs. Spatially, this state is exponentially localized at the interface, while in the spectrum of the system, it has a midgap position due to chiral symmetry. In our work, an interface supporting a topological edge state is created by repeating the weak bond twice. We apply local time-periodic perturbations associated with a single lattice site at the interface (site 0) by modulating the hopping amplitudes $J_{-1,0}(t)$ and $J_{0,1}(t)$ to its nearest neighbours and its local on-site potential $V_0(t)$. Since in the waveguide model the propagation distance $z$ plays the role of time [29], bending the $0^{th}$ waveguide sinusoidally with amplitude $A$, we implement such perturbations. Different frequency regimes are realized by varying the period $P$, while $A$ is always kept constant. Two different modulations are considered: in-plane (Fig. 1(a)) and out-of-plane direction (Fig. 1(b)). In contrast to previous studies [30, 31], we do not drive the bulk of the SSH model to guarantee that the topological invariants stay unchanged and the bulk gap stays open.

Topological invariants are global characteristics of bulk Hamiltonians. Thus, topological invariants of time-periodic systems must be obtained using the Floquet Hamiltonian if the bulk is periodically driven. However, in our case, the bulk is static. The topological invariant, *i.e.*, in our case, the winding number of the bulk, must not depend on the representation of our system; whether we use the Floquet picture or not it stays the same as in the static SSH model.

## II. RESULTS

### A. Theory

#### 1. *Floquet analysis*

We start with a theoretical analysis of our model based on the Floquet theory [32, 33] (see "Materials and methods"). Within this formalism, a band structure can be unambiguously described in terms of so-called quasienergies, analogues of the eigen energies in a time-independent problem. The corresponding Floquet states belong to the extended Hilbert space, which is a direct product of the usual Hilbert space and the space of periodic functions with period $P = 2\pi/\omega$. In the Floquet picture, our 1-dimensional time-periodic system can be displayed as a (1+1)-dimensional time-independent system [33]. Fig. 2 shows the static (1+1)D lattice, which is analogous to the SSH model with local harmonic perturbations of the topological defect at site $s$=0. It consists of an infinite number of SSH chains labelled by the Floquet index $n$ with the overall potential shifted by $-n\omega$ (throughout the paper, we set $\hbar = 1$). Periodic driving thus splits the band structure of the undriven system into infinitely many copies (Floquet replicas) spaced by $\omega$ [30,33]. Fig. 2 illustrates that local perturbations couple the chains only through the sites in the vicinity of the interface ($s$=-1,0,1) with the hopping amplitude $\Delta J/2$ due to variation in the couplings and $\Delta V/2$ due to on-site potential variations, both determined by the modulation amplitude $A$. Hence, by applying a local perturbation to the interface, we selectively populate the Floquet replicas of the topological edge state while the bulk states stay almost unaffected.

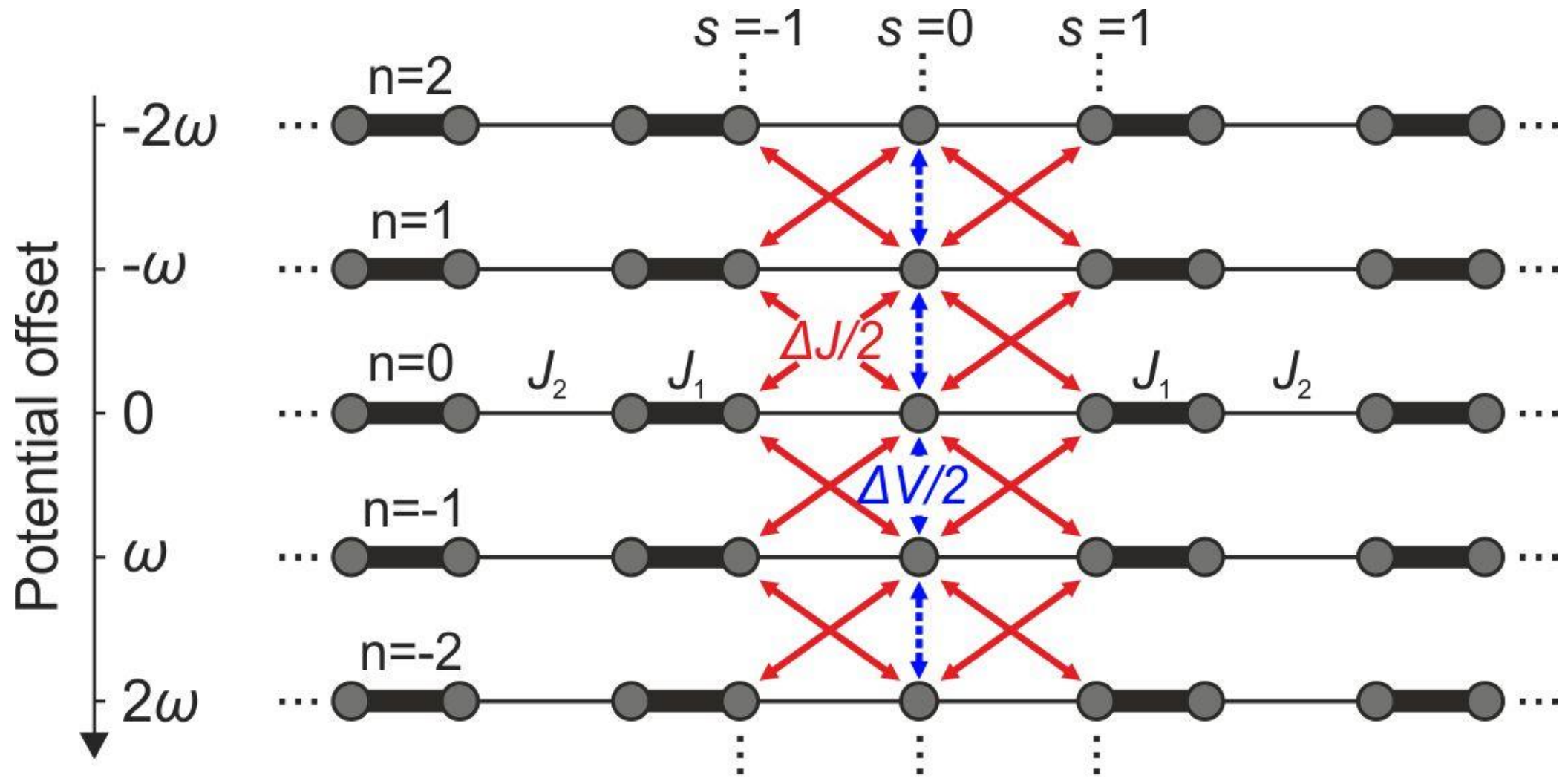

FIG. 2. (1+1)D time-independent analogue of the SSH model with local periodic perturbations at the interface and $J_1$ ($J_2$) being the large (small) hopping amplitude. The Floquet index $n$ enumerates coupled SSH chains, each with the overall potential $-n\omega$. Red and blue arrows denote the coupling between Floquet replicas ($n$) and sites ($s$) created by the harmonic driving of local couplings with amplitude $\Delta J$ and on-site potential with $\Delta V$ (compare with Eq. (12)).

In the following, we present results of the Floquet analysis for the model with the in-plane modulation of the topological defect. In this case, the couplings to the left $J_{-1,0}(t)$ and right $J_{0,1}(t)$ nearest neighbours of the $0^{\text{th}}$ site change with a phase shift of $\pi$. We choose $J_{-1,0}(t) = J_2 + \Delta J \sin(\omega t)$, $J_{0,1}(t) = J_2 - \Delta J \sin(\omega t)$, and $V_0(t) = 0$. As an initial condition, we solely excite the central lattice site $s$=0. The corresponding quasienergy spectrum is presented in Fig. 3 (a). Colour coding indicates the spectral weight of each Floquet state calculated using Eq. (15) in "Materials and methods".

As a reference, we consider the static system ($\Delta J$ = 0). In Fig. 3 (b), we plot the corresponding temporal evolution of the probability density $|\Psi(s,t)|^2$ (see Eq. (14), where $\Psi(s,t)$ is the projection of $|\Psi(t)\rangle$ on the lattice sites $s$ for the single-site input at the $0^{\text{th}}$ lattice site. Here, the excited bulk modes are spreading ballistically while the topological edge state shows itself as a fraction of the probability density localized at the interface. The momentum distribution of the probability density $|\widetilde{\Psi}(k,E)|^2$(see Fig. 3 (c)) features two cosine-shaped bands and a horizontal line in the middle of the band gap, a manifestation of the topological edge state.

In the low-frequency regime ($\omega < |J_1 - J_2|$), the first ($n = \pm 1$) replicas of the zero-energy mode lie inside the band gap (see the green arrows in Fig. 3 (e)). This is in full agreement with the edge-state counting rules of Floquet Hamiltonians [20]. We note that for all the modulation amplitudes accessible in the experiments, the effect of higher ($|n| > 1$) replicas is negligible (see the next subsection on the decay rates of a topological edge state). Fig. 3 (d) shows that $|\Psi(s,t)|^2$ stays localized at the $0^{\text{th}}$ lattice site.

This picture completely changes in the intermediate frequency regime ($|J_1 - J_2| < \omega < |J_1 + J_2|$), when the first replicas of the topological edge state enter the energy interval of the static bulk states inducing the aforementioned hybridization of bulk and edge states. As a result, the probability density delocalizes (Fig. 3 (f)), and the momentum distribution also shows the pronounced coupling, *i.e.*, the population of the zero-energy state drops drastically despite the non-trivial topological invariants (see the magenta arrow, Fig. 3 (g)), while the bulk bands gain more weight (green arrows in Fig. 3 (g)). No such coupling has been observed when driving the whole bulk of the system, as in [30]. There, the driving induces gaps to open when two Floquet replicas overlap, such that edge states are protected by the gaps from coupling to bulk states. Here, however, due to the spatially local driving, no such gaps are opened, and couplings can occur.

Finally, in the high frequency regime ($\omega > |J_1 + J_2|$), the $1^{\text{st}}$ Floquet replicas of the zero energy mode lie outside of the band, and no hybridization of bulk and edge states takes place. Consequently, the probability density is again localized, and the population of the topological edge state is restored (Figs. 3 (h), 3 (i)).

We note that in our system, no anomalous edge states [20] are created for any driving frequency since there is no periodic driving of the bulk. The periodic intensity modulation at the interface in Figs. 3 (d) and (h) results from beating of the topological edge state and its Floquet replicas. The asymmetry of the probability density distribution $|\Psi(s,t)|^2$ with respect to the interface in Figs. 3 (d), (f), and (h) results from the $\pi$ phase shift of the in-plane coupling modulation.

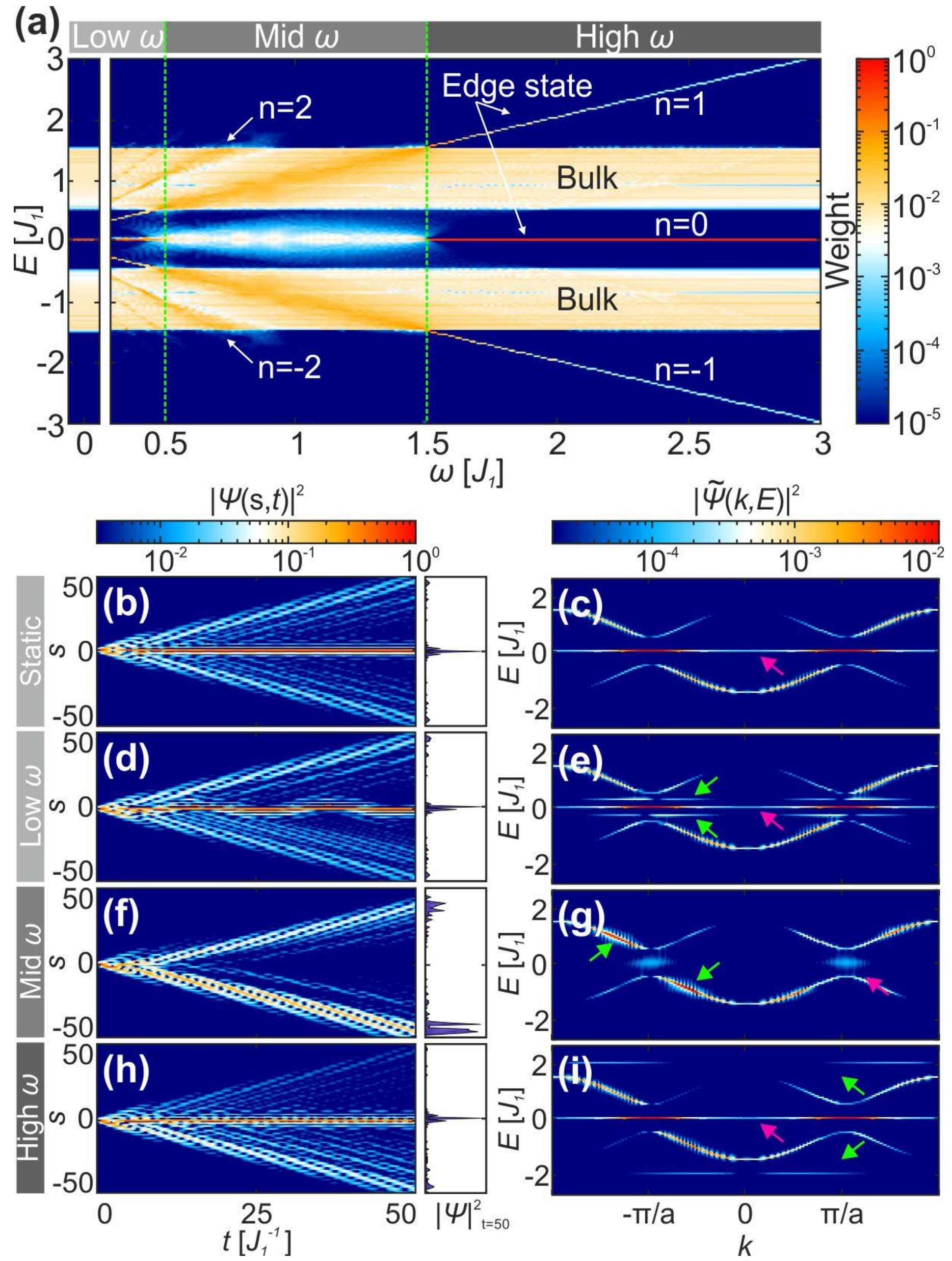

FIG. 3. (a) Frequency-dependent quasienergy spectrum with assigned weights in case of the in-plane perturbations at the interface. (b-i) Temporal evolution of the probability density (left) and corresponding momentum-resolved spectra (right) for (b, c) undriven case, (d, e) low frequency ($\omega$ = $0.3J_1$), (f, g) intermediate frequency ($\omega$ = $J_1$), and (h, i) high frequency ($\omega$ = $2J_1$). The histograms at the right side from (b), (d), (f), and (h) show the distribution of the probability density at $t = 50\, J_1^{-1}$. In the spectra, the magenta arrows point to the 0th Floquet replica of the edge state, while the green arrows indicate the locations of its 1st Floquet replicas. All the calculations were performed for the in-plane modulated SSH model with $2M$ = 100 dimers, $J_1$= 1, $J_2/J_1$= 0.5, and $\Delta J$ = $0.3J_1$. As initial conditions, we solely excited the 0th lattice site.

Analogous calculations for the out-of-plane perturbation (couplings are modulated in phase) show qualitatively the same behaviour. This case fulfils parity, which leads to a symmetric distribution of $|\Psi(s,t)|^2$ around the 0th site. Adding a periodic local on-site potential variation for the 0th site violates chiral symmetry by shifting the energy of the edge mode by the amount of $\Delta V$. However, this does not have a strong influence on the overall picture if the corresponding amplitude $\Delta V$ is smaller or on the order of $\Delta J$ (see supplementary material).

## 2. *Decay rate of a topological edge state*

Having the full time evolution given by Eq. (14) at hand, we can numerically calculate the decay rate of the topological edge state. Before addressing our calculation, we note that when lowering the frequency, the $n^{\text{th}}$ pair of replicas induces coupling between the edge state and the bulk in the frequency interval $\omega_n \in [|J_1 - J_2|/n,\ |J_1 + J_2|/n]$, which correspond to higher order transitions. However, the rates $\Gamma_n$ of the $n > 1$ order transitions are very low for realistic timescales (see [34]). For instance, the maximum $2^{\text{nd}}$-order transition rate is estimated to be two orders of magnitude smaller than the $1^{\text{st}}$. This is already far beyond the experimentally observable time scales, such that we can limit ourselves to the analysis of the $1^{\text{st}}$ order transition only. To calculate the decay rate, we use the eigenstate of the undriven model corresponding to zero energy as the initial condition $|\Psi(t=0)\rangle = |E=0\rangle$. The overlap of the resulting time-dependent solution $|\Psi(t)\rangle$ with $|E=0\rangle$ is then fitted by the following exponential function

$$|\langle E=0|\Psi(t)\rangle|^2 \approx (1-c)\exp(-\Gamma t) + c. \tag{1}$$

Here, the parameter $c$ is accountable for the value of this overlap at large times compared to the driving period, $c = |\langle E=0|\Psi(t \gg P)\rangle|^2$, while $\Gamma$ denotes the evolution rate. In Fig. 4 (a), the parameter $c$ is plotted versus the driving frequency $\omega$. If $\omega$ lies in the range of the bulk bands, the overlap (1) tends to zero with $t \to \infty$, signalling a complete depopulation of the topological edge state. Outside of the band, $c$ measures the population of the topologically protected edge state of the perturbed system. In case of small $\omega$, the parameter $c$ is strictly speaking not well-defined because the overlap (1) is oscillating at large times due to the uncertainty in the choice of the phase offset between the initial and final states. In the limit of low driving frequencies, the Floquet states are approximately given by the adiabatic eigenstates of the Hamiltonian, and for each point in time, the adiabatic eigenstates differ. This difference leads to the aforementioned uncertainty of the phase. We avoid this uncertainty by fixing the phase offset to equal integer multiples of 2π. In doing so, we obtain that $c$ approaches 1 when $\omega \to 0$ (complete localization of the light at the edge). At high frequencies, in contrast, the parameter $c$ becomes phase-independent and is uniquely determined for every $\omega$. In the limit $\omega \to \infty$, it again approaches 1.

The evolution rate $\Gamma$ determines the characteristic time scale at which the perturbed system decays from the given initial condition $|E=0\rangle$. In the intermediate frequency regime, where $c = 0$, $\Gamma$ plays the role of the decay rate of the topological edge state (blue line in Fig. 4 (b)). Fig. 4 (b) shows that the decay rates are largest around $\omega = 1J_1$, when the replicas are in the middle of the bulk band and the group velocity of bulk modes is largest.

The population decay of the edge state can easily be understood from the Floquet eigenvalue equation (13) and Fermi's golden rule arguments. When the first Floquet replica of the edge state becomes resonant with the bulk, the modulation perturbation ($H_{\pm1}$, see Eq. (12)) leads to hybridization with the continuum of bulk modes.

We compare the numerically determined decay rate $\Gamma$ with the transition rate $\Gamma_{\text{FGR}}$ calculated by Fermi's golden rule [35] (see the red line in Fig. 4 (b)). Both rates qualitatively follow the same trend. For driving frequencies close to $\omega = 0.5J_1$ and $\omega = 1.5J_1$ – *i.e.*, when the first Floquet replicas approach the borders of the bandgap – they coincide, while for frequencies around $\omega = 1J_1$, the rate $\Gamma_{\text{FGR}}$ is slightly larger than $\Gamma$. To understand this deviation better, we plot $\Gamma$ and $\Gamma_{\text{FGR}}$ at constant frequency $\omega = 1.01J_1$ in dependence on the driving amplitude $\Delta J$ (see Fig. 4 (c)). We find that in the perturbative regime of small driving amplitude $\Delta J$, both approaches coincide. With increasing $\Delta J$, the decay rate approaches the band gap energy, and $\Gamma$ deviates from $\Gamma_{\text{FGR}}$.

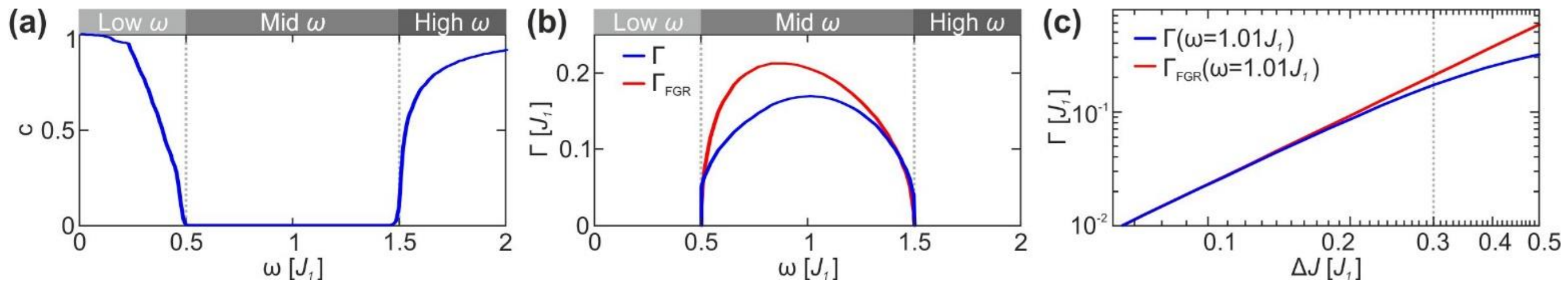

FIG. 4. (a) Fitting parameter $c = |\langle E=0|\Psi(t \gg P)\rangle|^2$ in dependence on the driving frequency $\omega$. (b) Decay rates $\Gamma$ of the topological edge mode in dependence on the driving frequencies in the intermediate frequency range, derived from the numerical solution (blue line) and Fermi's golden rule (red line), $\Delta J = 0.3J_1$. (c) The decay rate $\Gamma$ at a constant frequency $\omega = 1.01J_1$ versus the driving amplitude $\Delta J$. The dashed line highlights the experimentally relevant value of $\Delta J = 0.3J_1$.

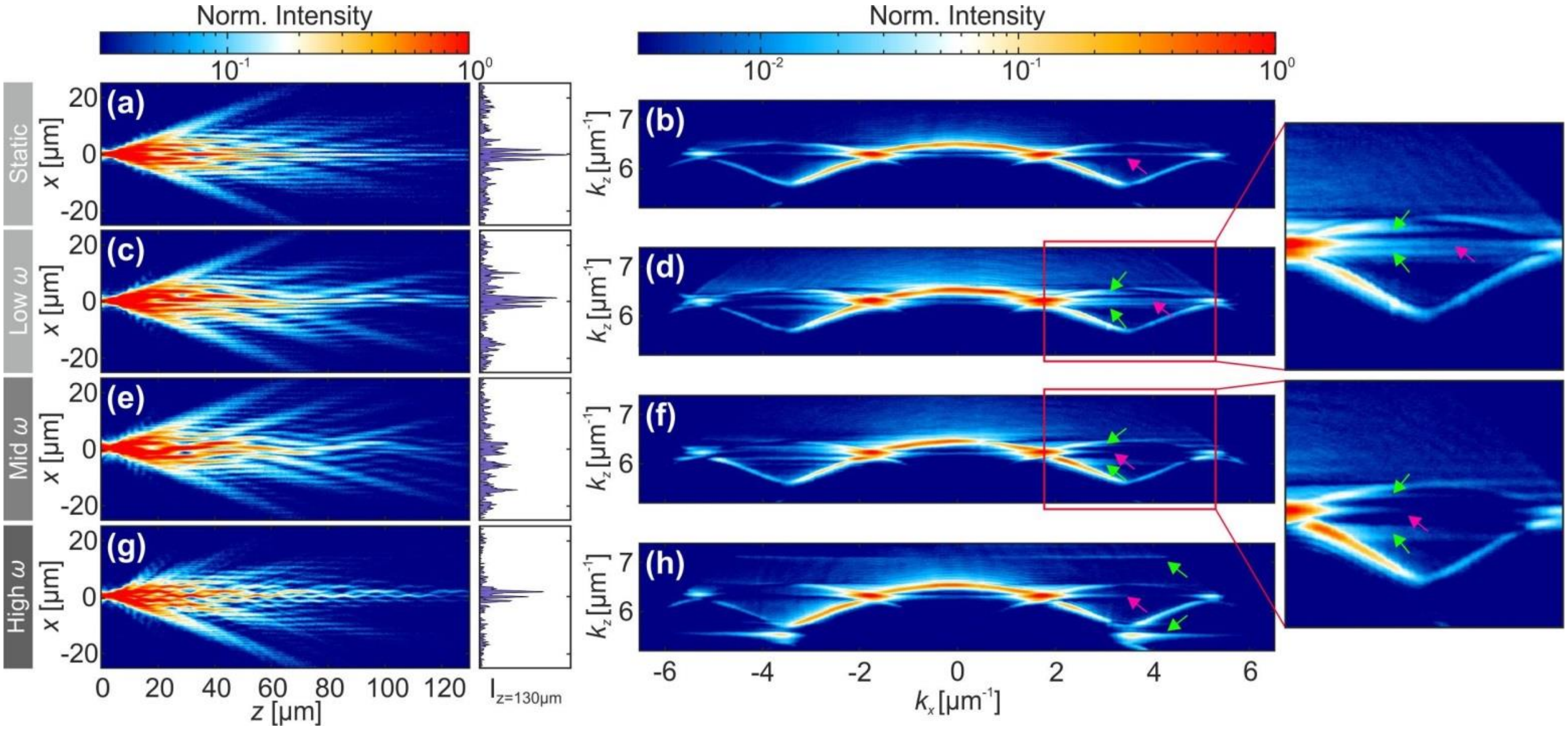

FIG. 5. Real- (left) and corresponding Fourier-space (right) leakage radiation micrographs of the DLSPPW arrays, analogous to the SSH model with a topological defect at $x$ = 0. The geometric parameters of all the arrays are chosen such that $J_2/J_1$= 0.5. (a) and (b) correspond to the static case. In (c-h), the defect is modulated in the in-plane direction ($\Delta J \approx 0.25 J_1$) with different frequencies: (c,d) low frequency regime ($\omega = 0.49 J_1$ corresponding to $P$=80 µm), (e,f) intermediate frequency regime ($\omega = 0.8 J_1$ corresponding to $P$=50 µm), (g,h) high frequency regime ($\omega = 4.9 J_1$ corresponding to $P$=8 µm). The histograms at the right side from the real-space images show the intensity distribution after the propagation distance of $z$=130 µm. In the Fourier-space images, the magenta arrows highlight the 0th Floquet replica of the edge state, while the green ones point to the locations of its first Floquet replicas.

## B. Experiments

We provide experimental evidence of the predicted effects using two photonic systems: arrays of dielectric-loaded surface plasmon-polariton waveguides (DLSPPWs) with in-plane modulation (Fig. 1(a)) and dielectric waveguide arrays with out-of-plane modulation (Fig. 1(b)). The technical aspects of these experiments are outlined in "Materials and methods".

We first consider in-plane modulation in DLSPPW arrays. In these experiments, leakage radiation microscopy gives direct access to the full real-space intensity distributions as well as the momentum resolved spectra in Fourier space (see Fig. 5). For all the measurements, surface plasmon polaritons (SPPs) were excited at a single waveguide in the centre of the array ($x$ = 0), which represents the interface. The geometric parameters of our samples are chosen such that $J_2/J_1 = 0.5$.

The case of the static SSH model [23] is shown in Figs. 5 (a) and 5 (b). In real space (Fig. 5 (a)), the excitation of the topologically protected mode results in localization of SPPs at the interface. The decaying intensity along the $z$-axis is due to radiation losses and absorption. However, this does not affect the topological properties of the system. The momentum- resolved spectrum of the static SSH model reveals the midgap position of this mode (see Fig. 5 (b)). We note that the asymmetry of the bulk bands arises from non-vanishing next-nearest neighbour coupling.

As predicted by Floquet theory, SPP localization at the interface in real space is also observed for modulation at low (Fig. 5 (c)) and high (Fig. 5 (g)) frequencies. In these cases (low and high frequencies), the Fourier-space measurements reveal that the 1st Floquet replicas do not overlap with the bulk bands; they either reside inside the band gap (Fig. 5 (d)) or outside of the bands (Fig. 5 (h)), respectively. In contrast, in the intermediate frequency regime ((Figs. 5 (e) and 5 (f))), the energy of the 1st Floquet replicas coincides with the static bulk states, and delocalization of SPPs into the bulk is observed (see the histogram in Fig. 5 (e)). Hence, we see clear experimental evidence of the depopulation of a topological edge mode by local driving in agreement with the results of the Floquet analysis discussed above.

Dielectric waveguide arrays are ideally suited for an out-of-plane modulation of the interface. In this set of experiments, we measure the intensity distribution at the output facet of the waveguide array. Fig. 6 (a) shows measurements for several structures with different periods (frequencies of modulation $\omega$) with otherwise identical parameters ($J_2/J_1$= 0.48) at a wavelength of $\lambda$=710 nm. Light is localized around the defect at the central site at $x$ = 0 for the low and high frequency regimes (topmost and bottom panels). In contrast, the light couples to the bulk modes for intermediate frequencies ($0.56 \leq \omega/J_1 \leq 1.12$).

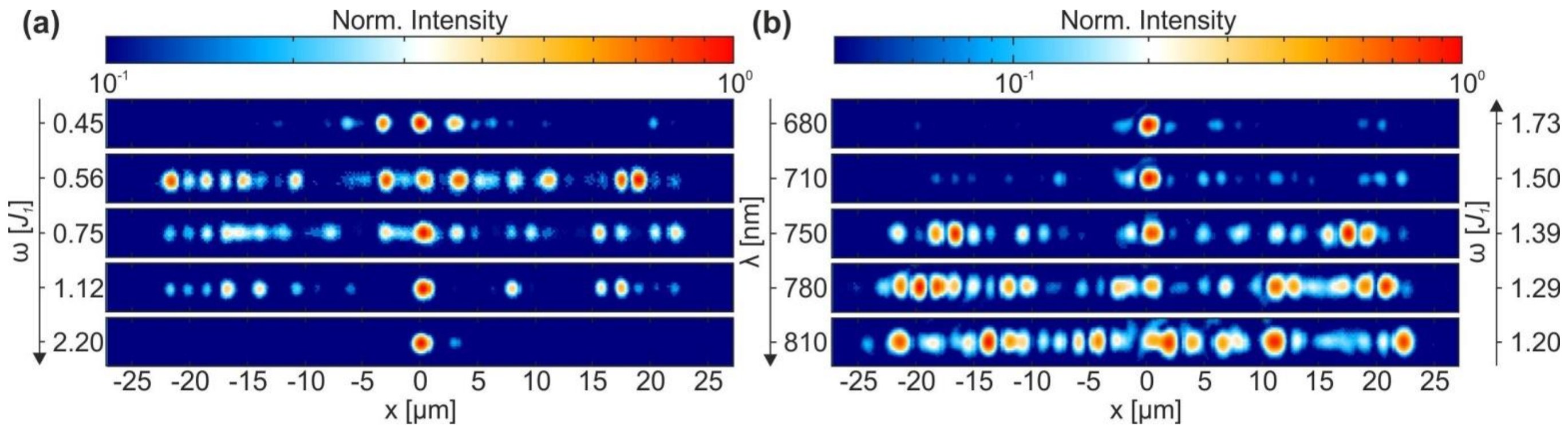

FIG. 6. Measurements in 3D printed dielectric waveguides for out-of-plane defect modulation. Shown are the intensities in the waveguides at the output facet. (a) Measurements for several structures with different periods (frequencies of modulation $\omega$) with otherwise the same parameters at fixed wavelength. For small frequencies, the light is localized around the defect (at $x = 0$). When the frequency is increased, light couples to the bulk states ($0.56 \leq \omega/J_1 \leq 1.12$) and localizes in the defect again for large frequencies. (b) In a structure with fixed period of the defect modulation, the wavelength is tuned. Light is delocalized, *i.e.*, couples strongly to the bulk, when the first Floquet mode hits the bulk band, starting at $\lambda = 750$ nm. Note that $J_2/J_1$ changes with the wavelength: $J_2/J_1 = 0.47$ (680 nm), 0.48 (710 nm), 0.52 (750 nm), 0.53 (780 nm), and 0.55 (810 nm).

To exclude any influence of fabricational deviations of distinct samples, the switching between different frequency regimes can also be done in one sample by changing the wavelength of the light (see Fig. 6 (b)). This changes the hoppings and therefore the ratio of $\omega/J_1$, the width of the band gap and the maximum energy of the bulk bands. Hence, the positions of the Floquet replicas relative to the bulk bands can be controlled. For a wavelength of 680 nm, the first Floquet replicas lie outside the bulk bands, corresponding to the high frequency regime. We see that the light is localized around the site at $x = 0$. With increasing wavelength, the energy of the replicas moves into the bulk band, and we again observe coupling to the bulk modes and spreading of the light, starting at a wavelength of 750 nm. This confirms that the observed effects are not due to fabricational deviations between different samples.

## III. DISCUSSION

In conclusion, we have shown that local driving of a defect in a system with non-trivial (bulk) topology can result in a depopulation of the edge state. The edge state energies are still symmetric with respect to zero, which in the fully static case, guarantees the energetic separation of the edge from the bulk states. The topological character of the bulk bands cannot be changed by any local perturbation, but nonetheless, we observe a dramatic change in the occupation and spectral characteristics of the edge state in certain frequency ranges, which can only be explained by hybridization with bulk states. This was demonstrated in calculations using Floquet theory and proven by measurements in plasmonic and dielectric waveguide arrays for in-plane and out-of-plane modulations of the defect. We moreover went beyond the qualitative picture by calculating the decay rates of the edge state. These calculations answer the question of how much our driven system deviates from the static one and how stable the edge state is. In the intermediate frequency range, enough energy is imparted to the system to destroy its topological protection, or, in more strict terms, the concept of topological protection is not valid any longer. In this paper, we set out to exactly demonstrate these limits.

Model systems as analysed here serve to control the localization and the steering of light via an external parameter. Our work gives insight into Floquet engineering of photonic systems and into the limited extent of topological protection in the periodically driven case.

# IV. MATERIALS AND METHODS

## A. Floquet analysis

### 1. *Introduction to Floquet theory*

Our theoretical analysis is based on the Floquet theory [32,33,36,37] that provides a general framework for treating systems governed by time-periodic Hamiltonians $\mathrm{H}(t+P) = \mathrm{H}(t)$ with a period $P = 2\pi/\omega$. According to this theory, a solution of the Schrödinger equation $\mathrm{i}\frac{\partial}{\partial t}|\psi(t)\rangle = \mathrm{H}(t)|\psi(t)\rangle$ can be written as a superposition of Floquet-states [33]

$$|\psi_\alpha(t)\rangle = \exp(-\mathrm{i}\varepsilon_\alpha t)\,|u_\alpha(t)\rangle, \tag{2}$$

where $\varepsilon_\alpha$ is the quasienergy and $|u_\alpha(t)\rangle$ is the associated Floquet mode. The quasienergies are defined up to integer multiples of $\omega$, and the Floquet modes are $P$-periodic functions $|u_\alpha(t+P)\rangle = |u_\alpha(t)\rangle$. The Floquet modes $|u_\alpha(t)\rangle$ thus belong to the extended Hilbert space, which is a direct product of the usual Hilbert space and the space of time-periodic functions with period $P = 2\pi/\omega$.

After the substitution of the Floquet ansatz (2) into the Schrödinger equation, we directly obtain an eigenvalue equation for $\varepsilon_\alpha$

$$\left(\mathrm{H}(t) - \mathrm{i}\frac{\partial}{\partial t}\right)|u_\alpha(t)\rangle = \varepsilon_\alpha|u_\alpha(t)\rangle. \tag{3}$$

Using spectral decomposition of the Hamiltonian and the Floquet modes

$$\begin{aligned}\mathrm{H}(t) &= \sum_{n=-\infty}^{\infty} \mathrm{e}^{-\mathrm{i}n\omega t}\,\mathrm{H}_n,\\ |u_\alpha(t)\rangle &= \sum_{n=-\infty}^{\infty} \mathrm{e}^{-\mathrm{i}n\omega t}\,|u_\alpha^n\rangle,\end{aligned} \tag{4}$$

we arrive at the time-independent Floquet equation

$$(\mathrm{H}_0 - n\omega)|u_\alpha^n\rangle + \sum_{m\neq 0} \mathrm{H}_m\,|u_\alpha^{n-m}\rangle = \varepsilon_\alpha|u_\alpha^n\rangle, \qquad \forall n \in \mathbb{Z}. \tag{5}$$

### 2. *Floquet analysis of the driven SSH model*

We now apply the Floquet approach to our system of interest. Let us first describe the corresponding Hamiltonian. We consider the systems sketched in Fig. 1, where the hopping amplitudes between the 0$^{\mathrm{th}}$ and $\pm 1^{\mathrm{st}}$ lattice sites $J_{-1,0}(t) = J_2 + \Delta J\sin(\omega t + \phi_1)$, $J_{0,1}(t) = J_2 + \Delta J\sin(\omega t + \phi_2)$ are time-dependent due to the modulation of the 0$^{\mathrm{th}}$ site, which also causes a small time-dependent on-site potential at site 0: $V_0(t) = -\Delta V + \Delta V\cos(\omega t)$. The phase factors are $\phi_1 = 0$, $\phi_2 = \pi$ for the in-plane modulation and $\phi_1 = \phi_2 = \pi/2$ for the out-of-plane modulation. Due to specific properties of each experimental realization, we can set $\Delta V = 0$ for the plasmonic waveguide model (Fig. 1 (a)), while for the dielectric waveguides (Fig. 1 (b)), $\Delta V \neq 0$ holds (see "Experimental methods" for details).

Assuming $4M$+1 lattice sites ($M$ dimers to either side of the defect and one unpaired site in the middle), the corresponding Hamiltonian can be written as a sum of time-independent and time-periodic parts

$$\mathrm{H}(t) = \mathrm{H}_0 + \mathrm{H}_\mathrm{P}(t), \tag{6}$$

where

$$
\begin{aligned}
\mathrm{H}_0 = & \sum_{s=-M+1}^{0} \left(J_1\, a_{2s-2}^{\dagger} a_{2s-1} + J_2\, a_{2s-1}^{\dagger} a_{2s}\right) \\
& + \sum_{s=0}^{M-1} \left(J_2\, a_{2s}^{\dagger} a_{2s+1} + J_1\, a_{2s+1}^{\dagger} a_{2s+2}\right) \\
& + \mathrm{h.c.} - \Delta V\, a_0^{\dagger} a_0
\end{aligned}
\tag{7}
$$

and

$$
\mathrm{H_P}(t) = \Delta J \sin(\omega t + \phi_1)\, a_0^{\dagger} a_{-1} + \Delta J \sin(\omega t + \phi_2)\, a_0^{\dagger} a_1 + \mathrm{h.c.} + \Delta V \cos(\omega t)\, a_0^{\dagger} a_0. \tag{8}
$$

We denote by $a_s^{\dagger}$ the creation operator acting at the lattice site $s$.

In the absence of the on-site potential offset ($\Delta V = 0$), the static Hamiltonian (7) as well as the time-dependent part (8) obey chiral symmetry. Indeed, if $\Delta V = 0$, the unitary and Hermitian operator

$$
\Gamma = \sum_{s=-M}^{M} a_{2s}^{\dagger}\, |0\rangle\langle 0|\, a_{2s} - \sum_{s=-M-1}^{M-1} a_{2s+1}^{\dagger}\, |0\rangle\langle 0|\, a_{2s+1} \tag{9}
$$

with $|0\rangle$ being the vacuum state, fulfils the relation $\Gamma \mathrm{H}_0 \Gamma^{\dagger} = -\mathrm{H}_0$. For the time periodic part, it holds

$$
\Gamma \mathrm{H_P}(t + t_0)\Gamma = -\mathrm{H_P}(-t + t_0), \tag{10}
$$

where $t_0 = P/4$ for the in-plane modulation and $t_0 = 0$ for the out-of-plane modulation, which implies chiral symmetry for Floquet systems (for a proof, see appendix A in [21]). Being chirally symmetric, our system possesses a zero-energy Floquet mode that exhibits a vanishing amplitude on every second lattice site [21, 31]. As was shown in [21], even a harmonic time-dependent on-site potential variation – while breaking chiral symmetry – does not affect the topological robustness of the system.

In our further calculations, we express $\mathrm{H}_0$ and $\mathrm{H_P}(t)$ as $(4M+1) \times (4M+1)$ matrices

$$
\mathrm{H}_0 = \begin{pmatrix}
\ddots & & & & & & & \\
 & & J_2 & & & & & \\
 & J_2 & 0 & J_1 & & & & \\
 & & J_1 & 0 & J_2 & 0 & & \\
 & & & J_2 & -\Delta V & J_2 & & \\
 & & & 0 & J_2 & 0 & J_1 & \\
 & & & & & J_1 & 0 & J_2 \\
 & & & & & & J_2 & \\
 & & & & & & & \ddots
\end{pmatrix}, \tag{11}
$$

and $\mathrm{H_P}(t) = \mathrm{H}_1 \mathrm{e}^{-\mathrm{i}\omega t} + \mathrm{H}_{-1} \mathrm{e}^{\mathrm{i}\omega t}$,

where the Fourier components $\mathrm{H}_{\pm 1}$ according to (4) are represented by

$$
\mathrm{H}_{\pm 1} = \mp \frac{1}{2} \cdot \begin{pmatrix}
\ddots & & & & \\
 & 0 & & & \\
 & & \begin{matrix} 0 & \mathrm{i}\Delta J \mathrm{e}^{\mp \mathrm{i}\phi_1} & 0 \\ \mathrm{i}\Delta J \mathrm{e}^{\mp \mathrm{i}\phi_1} & \mp \Delta V & \mathrm{i}\Delta J \mathrm{e}^{\mp \mathrm{i}\phi_2} \\ 0 & \mathrm{i}\Delta J \mathrm{e}^{\mp \mathrm{i}\phi_2} & 0 \end{matrix} & & \\
 & & & 0 & \\
 & & & & \ddots
\end{pmatrix}. \tag{12}
$$

The boxes highlight the central parts of the matrices, which are associated with the defect (0th lattice site in Fig. 1). Due to the spatially local character of perturbations of our model, all the elements outside of the box in the time-dependent part $\mathrm{H_P}(t)$ are zero.

The Floquet equation (5) can be represented as the following eigenvalue problem with an infinite block-matrix operator

$$
\begin{pmatrix}
\ddots & & & & \\
\mathrm{H}_1 & \mathrm{H}_0 + \omega \mathbb{I} & \mathrm{H}_{-1} & & \\
 & \mathrm{H}_1 & \mathrm{H}_0 & \mathrm{H}_{-1} & \\
 & & \mathrm{H}_1 & \mathrm{H}_0 - \omega \mathbb{I} & \mathrm{H}_{-1} \\
 & & & & \ddots
\end{pmatrix}
\begin{pmatrix} \vdots \\ u_\alpha^{-1} \\ u_\alpha^{0} \\ u_\alpha^{+1} \\ \vdots \end{pmatrix}
= \varepsilon_\alpha
\begin{pmatrix} \vdots \\ u_\alpha^{-1} \\ u_\alpha^{0} \\ u_\alpha^{+1} \\ \vdots \end{pmatrix}. \tag{13}
$$

Here, the index of the operator elements runs over the lattice sites. This equation reveals an illustrative interpretation of the Floquet approach; it transforms our 1D time-periodic problem into a (1+1)D time-independent one with the Floquet replicas building up the synthetic dimension [11, 13, 30]. Eqs. (7)-(13) for the SSH model with local driving are summarized pictorially in Fig. 2 on the (1+1)D lattice. This lattice consists of an infinite number of SSH chains labelled by the Floquet index $n$ with the overall potential shifted by $-n\omega$. Each lattice site can be now identified by two numbers $[n,s]$, where $s$ is the site index within each chain and $n$ labels the Floquet replicas of the system [30]. Due to local perturbations, the chains are coupled to each other only through the sites in the vicinity of the topological defect ($s = -1,0,1$). The harmonic variation of the hoppings $J_{-1,0}(t)$ and $J_{0,1}(t)$ thus induces the bonds between the sites $[n,0]$ and $[n \pm 1,\pm 1]$, $\forall n$ with the hopping amplitude $\Delta J/2$. Likewise, a harmonic on-site potential variation at the $0^{\mathrm{th}}$ lattice site with the amplitude $\Delta V$ creates bonds between the central sites $[n,0]$ and $[n \pm 1,0]$ $\forall n$ with the hopping term $\Delta V/2$.

The quasienergy spectrum of the periodically driven system consists of infinitely many copies of the spectra of the undriven system spaced by $\omega$ [30, 33]. In the Floquet picture, the energy of such a Floquet replica of the edge state $(n\omega)$ can have the same value as that of a bulk state $\varepsilon_\alpha = n\omega$. When edge and bulk states hybridize, the edge state depopulates into the bulk due to the local time-periodic coupling.

A sufficiently large truncated version of equation (13) yields eigenvectors and eigenvalues that converge well. We restrict ourselves to the quasienergies from the first Floquet Brillouin zone $\varepsilon \in [-\omega/2, \omega/2[$. The corresponding eigenvectors contain the Fourier components of the Floquet modes $|u_\alpha^n\rangle$, where each of them is associated with the energy $\varepsilon_\alpha^n = \varepsilon_\alpha + n\omega$. The complete solution of the Schrödinger equation is given by

$$|\Psi(t)\rangle = \sum_\alpha C_\alpha \sum_n \exp(-\mathrm{i}\varepsilon_\alpha^n t)\, |u_\alpha^n\rangle, \tag{14}$$

where the constants $C_\alpha = \langle u_\alpha(0)|\Psi(0)\rangle$ are determined by the initial condition $|\Psi(0)\rangle$. The temporal Fourier transform of the wave function (14) reads $|\psi(E)\rangle = \sum_{\alpha,n} C_\alpha\, |u_\alpha^n\rangle \delta(E - \varepsilon_\alpha^n)$ and motivates defining the spectral weight at energy $E = \varepsilon_\alpha^n$ by

$$w(\varepsilon_\alpha^n) = |C_\alpha|^2 \langle u_\alpha^n | u_\alpha^n \rangle. \tag{15}$$

The sum over all weights is normalized to one. Note that $|\Psi(t)\rangle$ is a time-dependent vector whose components, corresponding to different lattice sites $s$, take the value of a wave function $\Psi(s,t)$. The 2D Fourier transform $\tilde{\Psi}(k,E)$ yields the momentum representation of the wave function $\Psi(s,t)$.

## B. Experimental methods

### *1. Dielectric-loaded surface plasmon-polariton waveguides*

The DLSPPW arrays were fabricated by negative-tone grey-scale electron beam lithography [24]. Figure 7 (a) depicts an electron micrograph of a typical sample. The DLSPPWs consist of poly(methyl methacrylate) (PMMA) ridges deposited on top of a 60 nm thick gold film evaporated on a glass substrate. Additionally, 5 nm of Cr was used as an adhesion layer. The width and the height of each waveguide were designed to be 250 nm and 110 nm, respectively, to guarantee single-mode operation at the working light wavelength of $\lambda$=980 nm. To keep the heights of the waveguides constant, the proximity effect in the lithographic process was compensated by equalizing the background dose. The waveguide geometry was controlled after fabrication by atomic force microscopy. In all the samples, the short distance was $d_1$=0.7 µm, and the long distance was $d_2$=1.1 µm. These separations correspond to coupling constants $J_1$=0.16 µm$^{-1}$ and $J_2$=0.08 µm$^{-1}$, respectively. The propagation constant of a single DLSPPW is $\beta$=6.65 µm$^{-1}$. These parameters were chosen to ensure sufficient coupling between the adjacent waveguides and to introduce perceptible dimerization to see topological effects. The position of the central waveguide was modulated sinusoidally, resulting in

$$J_{0,1}(t) = J_1 \cdot p_1 \exp(-p_2 \cdot A \sin(\omega t)), \tag{16}$$

where $p_1$=0.49 and $p_2$=1.75 µm$^{-1}$ are fitting parameters and $\omega$ is the modulation frequency. For all the samples, the maximum deflection of the central waveguide was chosen to be $A$=0.3 µm, being a good trade-off between bending losses and the strength of dynamic effects. It corresponds to the coupling variation of $\Delta J \approx 0.25 J_1$ (for linear approximation of the exponent in (16)). Varying the period $P$ from 8 µm up to 80 µm, we realized different frequency regimes. Due to strong confinement of the SPPs, we can neglect the variation of the effective refractive index due to curvature of the waveguide, *i.e.*, we can set the on-site potential $V_0(t) \approx 0$.

SPPs were excited by focusing a TM-polarized laser beam (the (numerical aperture (NA) of the focusing objective is 0.4) onto the grating coupler (see the red dotted box in Fig. 7 (a)), which was fabricated on top of the central waveguide. The propagation of SPPs in the array was monitored by real- and Fourier-space leakage radiation microscopy (LRM) [25, 38]. The leakage radiation as well as the transmitted laser beam were both collected by a high

NA oil immersion objective (Nikon 1.4 NA, 60x Plan-Apo). The transmitted laser was filtered out by placing a knife edge at the intermediate back focal plane (BFP) of the oil immersion objective. The remaining radiation was imaged onto an sCMOS camera (AndorZyla). Real-space SPP intensity distributions were recorded at the real image plane, while the momentum-space intensity distribution was obtained by imaging the BFP of the oil immersion objective.

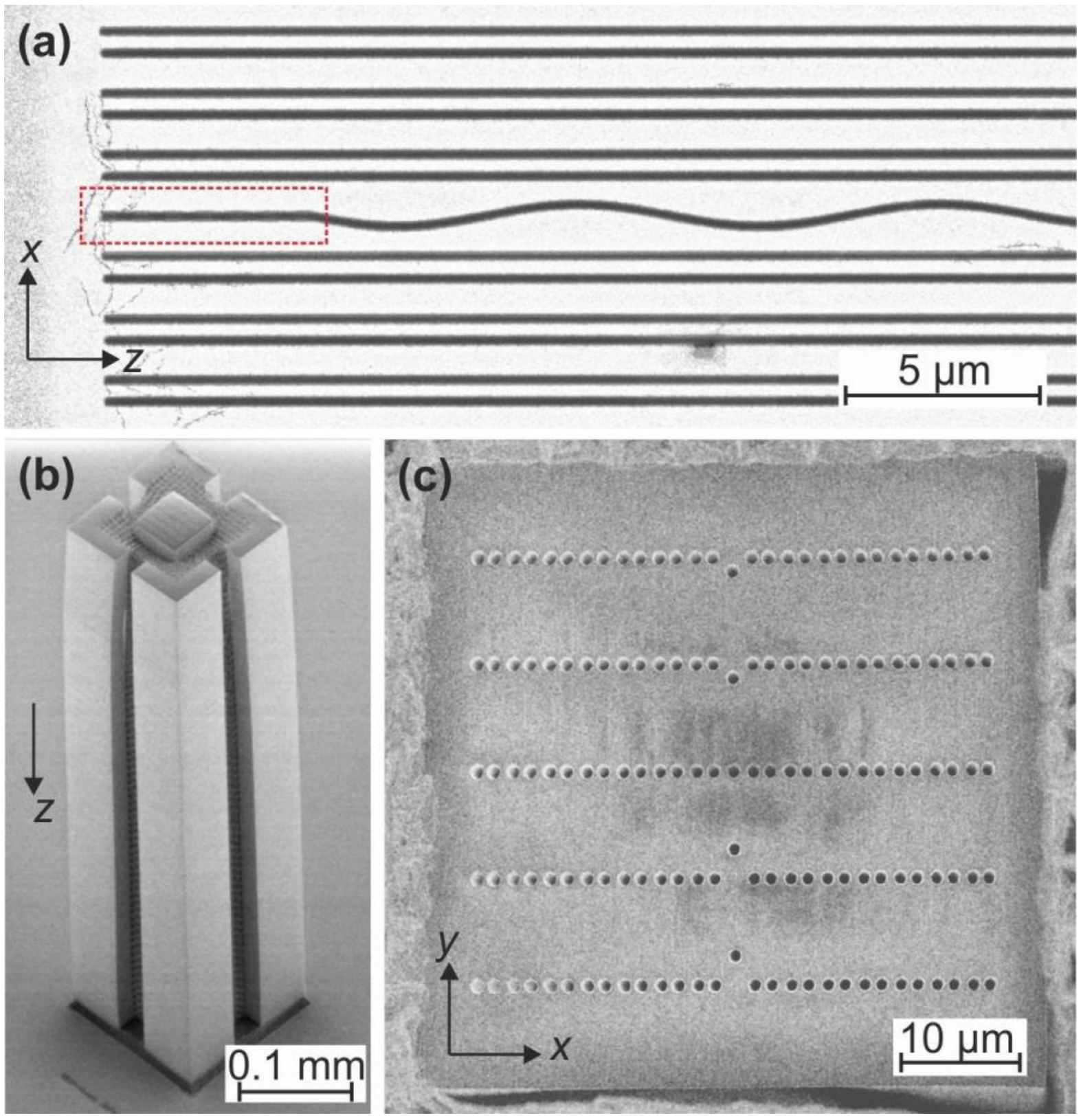

FIG. 7. Scanning electron micrographs of the plasmonic (a) and dielectric (b: side, c: top) waveguide samples. In (a) $P$=10 μm, and the red dotted box highlights the grating coupler. The sample shown in (b) and (c) corresponds to 5 arrays with different periods of defect modulation.

### *2. Dielectric waveguides*

Dielectric waveguide arrays were fabricated by direct 3D laser writing. Side and top views of one dielectric waveguide sample are shown in Figs. 7 (b) and (c), respectively. The sample fabrication included two steps [26]. First, the inverse of the waveguide structure was 3D-printed by two-photon lithography in a negative tone photoresist (IP-Dip, Nanoscribe). After development, the hollow structure was then infiltrated with SU8-2 (MicroChem) to create the waveguides. Baking the sample on a hotplate at 150˚C for 3 minutes, after ramping up the temperature at 10˚C per minute, the SU8 was solidified. The resulting refractive indices of the outside material and the waveguide core were $n_0$=1.54 and $n_{core}$=1.59, respectively. The radius of the waveguides $r$ as well as the small distance $d_1$ and large distance $d_2$ were measured by scanning electron microscopy. For all the samples, we fixed these parameters to be $r$=(0.52 ± 0.03) μm, $d_1$=(1.42 ± 0.02) μm and $d_2$=(1.69 ± 0.01) μm. For out-of-plane modulation of the defect, the couplings from site 0 to its left and right neighbours are equal, $J_{-1,0} = J_{0,1}$. $J_{0,1}$ scales exponentially as

$$J_{0,1}(t) \propto \exp\left(-p\sqrt{d_2^2 + \frac{A^2}{2}\left(1-\cos\left(\omega t\right)\right)}\right) \qquad (17)$$

The parameter $p$ depends on the refractive index contrast, used wavelength, etc.; $A$ is the maximum deflection of the waveguide, and $\omega$ is the frequency of the modulation. In the experiments presented in Fig. 6 (a), $J_{0,1}$ varied from $0.48J_1$ to $0.13J_1$, while for those in Fig. 6 (b), the variation depended on the wavelength: from $0.47J_1$ to $0.01J_1$ (680 nm), from

0.48$J_1$ to 0.01$J_1$ (710 nm), from 0.52$J_1$ to 0.02$J_1$ (750 nm), from 0.53$J_1$ to 0.02$J_1$ (780 nm) and from 0.55$J_1$ to 0.03$J_1$ (810 nm).

In the dielectric waveguides, we also have to take into account an additional local on-site potential at site 0 of

$$V_0(t) = -\Delta V + \Delta V \cos(\omega t) \quad (18)$$

This is because one can rewrite a curved waveguide in terms of a straight waveguide with changed refractive index [39]. We estimated the amplitude $\Delta V$ to be proportional to

$$\Delta V = 2r\, n_{\mathrm{core}}\, A\, (\omega/2)^2 \pi/\lambda \quad (19)$$

with the waveguide diameter $2r$. This additional local on-site potential at site 0 shifts the energy of the edge state by the amount of $\Delta V$. As there is no difference if the waveguide bends up or down, the on-site potential and couplings vary with twice the waveguide period. Therefore, we define the period of modulation $P$ to be half the waveguide period (see Fig. 1(b) bottom).

As shown in Fig. 7 (c), five arrays with defects with different periods $P$ ($2P$=(979 ± 14) µm, (783 ± 11) µm, (588 ± 11) µm, (392 ± 6) µm, (200 ± 3) µm) were fabricated in one sample. The amplitude of modulation was fixed to be $A$=(1.36 ± 0.04) µm. A different sample was used for the measurements shown in Fig. 6 (b). Here, $A$=(2.63 ± 0.08) µm, and $2P$=(302 ± 2) µm.

To conduct the measurements, the beam from a tuneable white light laser (SuperK EVO, NKT photonics) was sent through a VARIA (NKT photonics) filter box to select a certain wavelength (bandwidth 10 nm). The beam was then expanded and focused through an objective (Zeiss, NA 0.4, 20x) into the defect waveguide at site 0 at the input facet. We observed the intensity distribution in the sample at the (opposite) output facet by imaging it through an identical objective and a lens onto a CMOS-camera (Thorlabs). This corresponds to a propagation of 833 µm in $z$ or approximately 24 hops with $J_1$.

## ACKNOWLEDGEMENTS

FL is supported by a fellowship through the Excellence Initiative MAINZ (DFG/GSC 266). We acknowledge support by the Nano Structuring Center Kaiserslautern and by the Deutsche Forschungsgemeinschaft through CRC/Transregio 185 OSCAR. We thank Axel Pelster for theoretical discussions. We also thank Mark Kremer, Gerard Queralto and Moshe-Ishay Cohen for interesting discussions regarding the interpretation of our results.

## CONFLICT OF INTERESTS

The authors declare that they have no conflict of interest.

## AUTHOR CONTRIBUTIONS STATEMENT

Z.C. and C.J. contributed equally to this work.

Z.C. fabricated the plasmonic samples, conducted the corresponding optical experiments, and performed the Floquet analysis. S.L. conceived the plasmonic experiment. C.J. conceived together with G.v.F. the dielectric experiment, fabricated the dielectric waveguide samples, performed the corresponding measurements and contributed to the development of the theoretical understanding of the effects. Z.C. and C.J. wrote the first draft of the paper. C.D. developed together with S.E. the Floquet analysis and contributed to the theoretical understanding of the effects. M.F. and F.L. contributed to the theoretical understanding of the effects. All the authors discussed the results and reviewed the manuscript.

# Supplementary material: Limits of topological protection under local periodic driving

Z. Cherpakova,[1,*] C. Jörg,[2,*] C. Dauer,[2] F. Letscher,[2,3] M.Fleischhauer,[2] S. Eggert,[2] S. Linden,[1,+] and G. von Freymann[2,4,±]

[1]PhysikalischesInstitut, Universität Bonn, 53115 Bonn, Germany

[2]Physics Department and Research Center OPTIMAS, TU Kaiserslautern, 67663 Kaiserslautern, Germany

[3]Graduate School Materials Science in Mainz, 67663 Kaiserslautern, Germany

[4]Fraunhofer Institute for Industrial Mathematics ITWM, 67663 Kaiserslautern, Germany

*these authors contributed equally to this work

+ linden@physik.uni-bonn.de

± georg.freymann@physik.uni-kl.de

**Role of the on-site potential for the experiments:**

In case of the DLSPPW waveguides the on-site potential is very small due to strong confinement of SPPs. We have performed the measurements with isolated curved waveguides and estimated $\Delta V$ to be approximately three times less than the amplitude of coupling variation $\Delta J$ even at the highest frequency. Our Fourier-space measurements also confirm that (see Fig. 5 (h)).

In the case of the dielectric waveguides, we have calculated the band structure for the parameters used for the experiments in Fig. 6(a). The result of the calculations is shown below in Fig. S1. It is seen that the only visible effect of the on-site potential variation caused by the curvature of the waveguides is the overall shift of an edge state together with its replicas towards higher energies by $\Delta V$. This effect is more pronounced for high frequencies, as $\Delta V$ scales with $\omega^2$ (see Eq. 19). Due to this shift, one of the replicas hits the bulk bands a bit earlier than the other with increasing frequency. For $\omega/J_1<1.6$ we have $\Delta V<\Delta J$. However, even for $\omega/J_1=2.2$, $\Delta V$ is still smaller than the bandgap, and $\omega$ is so large that the first Floquet replicas lie well outside the bulk bands. For the measurements in Fig. 6(b), $\Delta V<\Delta J$ is valid for all wavelengths used. Therefore, the on-site potential does not change the overall picture qualitatively.

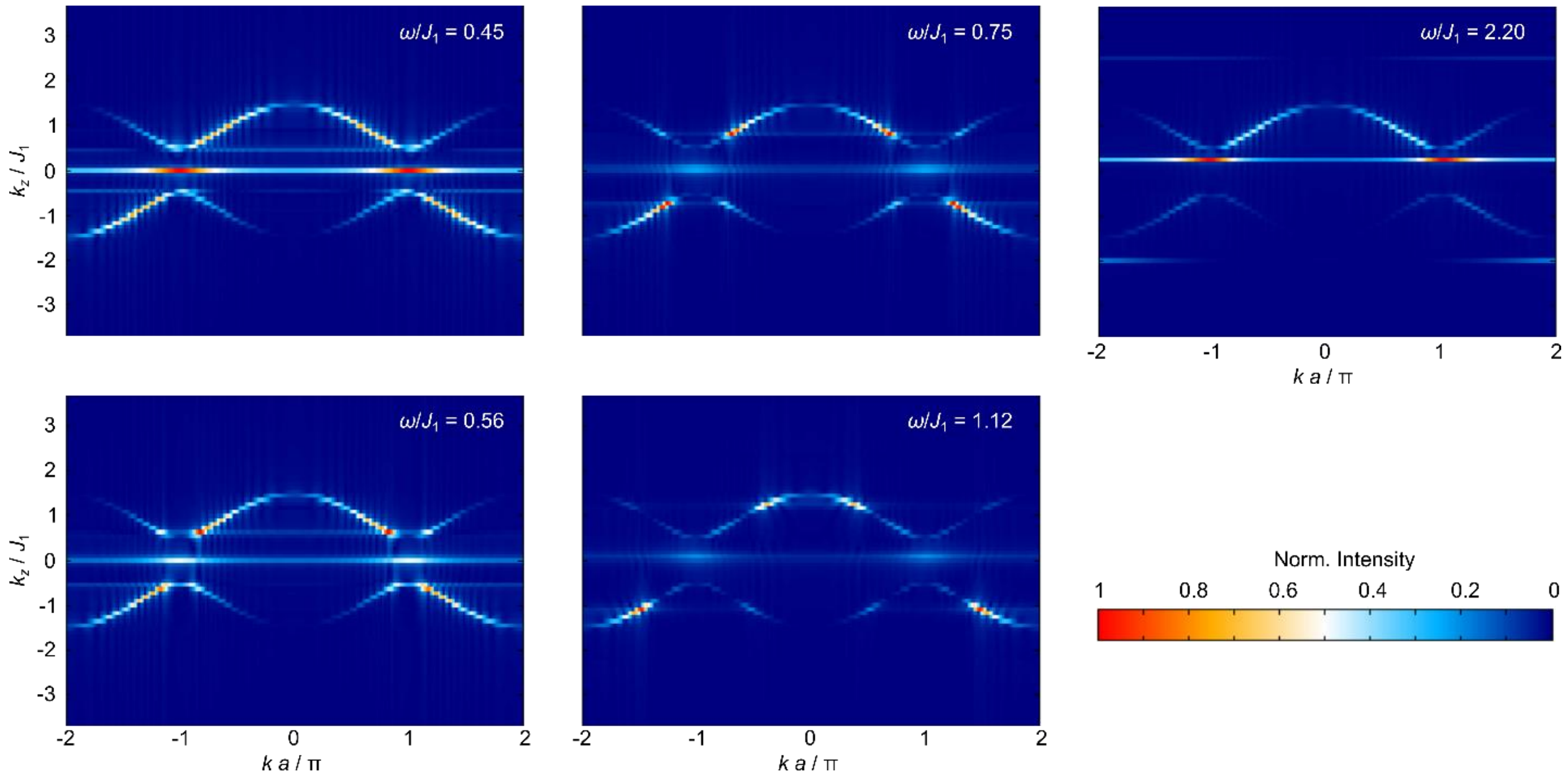

Fig. S1: Numerically calculated band structures for the experimental parameters of Fig. 6(a).